# Internet search effort on Covid-19 and the underlying public interventions and epidemiological status

**Aristides Moustakas[1],[*]**

1. Natural History Museum of Crete, University of Crete, Greece

* Corresponding author

Email: arismoustakas@gmail.com

**Abstract**

Disease spread is a complex phenomenon requiring an interdisciplinary approach. Covid-19 exhibited a global spatial spread in a very short time frame resulting in a global pandemic. Data of web search effort in Greece on Covid-19 as a topic for one year on a weekly temporal scale were analyzed using governmental intervention measures such a s school closures, movement restrictions, national and international travelling restrictions, stay at home requirements, mask requirements, financial support measures, and epidemiological variables such as new cases and new deaths as potential explanatory covariates. The relationship between web search effort on Covid-19 and the 16 in total explanatory covariates was analyzed with machine learning. Web search in time was compared with the corresponding epidemiological situation, expressed by the $R_t$ at the same week. Results indicated that the trained model exhibited a fit of $R^2$ = 91% between the actual and predicted web search effort. The top five variables for predicting web search effort were new deaths, the opening of international borders to non-Greek nationals, new cases, testing policy, and restrictions in internal movements. Web search peaked during the same weeks that the $R_t$ was peaking although new deaths or new cases were not peaking during those dates, and $R_t$ rarely is reported in public media. As both web search effort and $R_t$ peaked during 1-15 August 2020, the peak of the tourist season, the implications of this are discussed.

**Introduction**

Disease spread is a complex phenomenon and requires an interdisciplinary approach spanning from medicine to statistics and social sciences (Christakos et al. 2006). Covid-19 exhibited a global spatial spread in a relatively very short time frame resulting in being characterized as a pandemic by the World Health Organisation (WHO 2020). Quantifying human psychology or motivation to be informed about facts is admittedly a complex and potentially polarized issue (Kreps and Kriner 2020). Health-related issues are particularly difficult, as humans in an effort to protect their own or other peoples' lives may need to compromise their daily habits, personal and social freedom or financial interests (Clinton et al. 2021). Public interest in a disease spread is a dynamical phenomenon, as different factors may dominate depending on the nature of a disease or the economic status, or the health system of the country they reside and these factors exhibit strong interaction effects (Green et al. 2020). In addition, intervention measures employed in order to control the disease are dynamically evaluated by the public (Bento et al. 2020).

Quantifying Covid-19 web search interest into quantifiable characteristics in a consistent and reproducible way, could facilitate explaining human preferences as well as divergences between intervention measures (Bento et al. 2020, Du et al. 2020, Zitting et al. 2021). In addition, it can provide a way to understand mass psychology during such events (Du et al. 2020, Zitting et al. 2021). To date there is no known anti-viral treatment, and vaccination against Covid-19 was implemented relatively recently (Molina et al. 2020). Therefore the available options against the virus were the immune system and health status of each individual, the social behaviour, restrictions of movement or public events, and testing (Mack et al. 2007, Guo et al. 2020, Utsunomiya et al. 2020). Thus, there were relatively few options to be employed in order to diminish the spread of the disease (Haug et al. 2020).

Substantial analysis has been conducted regarding the efficacy of different intervention measures in the disease control (Haug et al. 2020). In addition, the impacts of social distancing and movement restrictions have been reported from a social and psychological perspective. However, it is hard to quantify what is the public risk perception and what is triggering the highest public attention around Covid-19 (Cori et al. 2020, Dryhurst et al. 2020). Public attention could be triggered by fear of getting sick or fear of death (Menzies and Menzies 2020, Pradhan et al. 2020). However, public interest could also be driven by fear of poverty, unemployment, or financial insecurity, or reciprocally financial aid or dept relief (Giordano 2020, Mamun and Ullah 2020, Esobi et al. 2021). Additional causes of public interest are disruption of individual and social freedom of movement and gathering, outgoing and entertainment, national and international travels, and school closures (Clinton et al. 2021). There clearly is a trade-off between freedom, safety, democracy, and economy (Landman and Splendore 2020, Simon 2020). Quantifying the public interest associated with the intervention measures comparable with the corresponding epidemiological parameters and ranking the relative contribution of each factor on the web search interest regarding Covid-19 may facilitate understanding elements of human nature.

The potential societal impact of interventions may become clearer when the contribution of these factors into new Covid-19 web search effort are ranked accounting for the contribution of each factor (Haug et al. 2020) to the total web search interest. In addition, assuming a good model fit, a data-trained predictive model could provide further insights into societal impacts of future interventions (Moustakas 2018, Lampos et al. 2021). In this study data from web search effort in Covid-19 in Greece were employed since the emergence of the first case of Covid-19 in the country. The web search effort was used as a proxy of societal interest in the topic (Effenberger et al. 2020, Szmuda et al. 2020) and the corresponding epidemiological, social distancing, and economic data were analyzed using machine learning (Demertzis et al. 2020). The study sought to quantify the predictive accuracy of a trained ML model with web search effort as a dependent variable validated against data. Feature importance of each epidemiological, social distancing and economic explanatory variable on Covid-19 web search effort was quantified. Finally the web search effort was compared against the epidemiological situation in the country.

**Methods**

*Web search data*

Data mining (Moustakas and Katsanevakis 2018) from Google Trends (https://trends.google.com/trends/)  was performed in order to connect these data with the current status at the time of search - see e.g. (Alicino et al. 2015, Effenberger et al. 2020, Szmuda et al. 2020). Google Trends is a publicly available repository of information on real-time web user search patterns of individuals that use Google as their search engine. Google trends was used it to assess the public interest in Greece in terms of web search interest of the Covid-19 topic as time series . In order to do this, the use of 'Coronavirus disease 2019' as a topic in Google trends from 26 February 2020  to 14 February 2021 was analyzed. Throughout the analysis week 1 (first week) is the week 24 Feb 2020 - 1 March 2020, while week 51 (last week) is the week 8 Feb 2021 - 14 Feb 2021. The use of ' Coronavirus disease 2019' as a topic (thereby Covid-19)  and not as a simple term/word ensures that the results exclude those searches with the word 'Covid-19' but referring to other different contexts e.g. other Corona viruses or songs containing the term 'Covid' in their title or lyrics etc. Contrastingly, synonyms such as 'Covid19' or 'Corona-19' are also included when considering the Covid-19 as a topic, allowing a more comprehensive inclusion of searches. Regarding languages, our analysis included web searches in all major languages (and not just Greek) covered by Google translator provided that the search on the topic was conducted from a Greek IP address or internet provider.

In order to account for the fact that the total number of users and thus the number of searches are differing over time regardless of the search term, web search data are normalized by Google Trends so as to represent search interest relative to the highest search keyword in Greece for that date(Burivalova et al. 2018). For example a value of Google search on Covid-19 of 100% means that the topic 'Covid-19 was the most popular search term on that date. A value of 50 means that the term is half as popular than the most popular searched topic on that date.  A score of 0 means that there was not enough data for the topic on that date (Burivalova et al. 2018). Google translator facilitated searching for the web search effort of the search topic of Covid-19 across all major languages, for searches conducted via a Greek IP. The temporal resolution of the data is one week.

*Epidemiological data*

New confirmed Covid-19 cases (*New cases*) per day in Greece were used a proxy of disease spread as a scale variable. New cases and not new cases per million or new cases smoothed per week is the feature of Covid-19 spread commonly reported in the media. Similarly, new confirmed Covid-19 deaths (*New deaths*) per day (scale variable) were used as a proxy of mortality, as this is the mortality related variable reported in the news and non-scientific reports and outputs. New Covid-19 deaths are deaths of people that were Covid-19 infected. The original temporal resolution of the data was one day. As the temporal resolution of the web search effort data is one week, new cases and new deaths were averaged per week and weekly average values were employed throughout the analysis. Data regarding new Covid-19 cases and deaths per day were retrieved from the database 'Our world in data' (Roser et al. 2020) publicly available at: (Our_World_in_Data 2021b). The original data derive from the European Centre for Disease Prevention and Control (ECDC), an EU agency with the aim to strengthen Europe's defense against infectious diseases from 26/02/2020 until 14/12/2020 when ECDC switched from daily data to weekly data, and from 15/12/2020 till 14/02/2021 from John Hopkins University that reports data daily (Our_World_in_Data 2021b). New cases and new deaths were averaged per week to match the temporal resolution and dates of the web search effort data.

*Public intervention data*

All data were recorded at a daily temporal resolution as time series as ordinal factor variables with different factor levels (unless otherwise explicitly stated) and averaged per week to match the resolution and dates of web search effort weekly data. These are publicly available from the database 'Our world in data' (Roser et al. 2020) at: (Our_World_in_Data 2021a)

*Containment and closure policies (C)*

School closure policy (*C1_School closing*) was recorded with levels:  0 - no measures, 1 - recommend closing or all schools open with alterations resulting in significant differences compared to non-Covid-19 operations, 2 - require closing (only some levels or categories, eg just high school, or just public schools, 3 - require closing all levels.
Workplace closure policy (*C2_Workplace closing*) with levels: 0 - no measures, 1 - recommend closing (or recommend work from home), 2 - require closing (or work from home) for some sectors or categories of workers, 3 - require closing (or work from home) for all-but-essential workplaces (eg grocery stores, doctors).
Cancelation of public events (*C3_Cancel public events*) with levels: 0 - no measures, 1 - recommend cancelling, 2 - require cancelling.
Gathering restrictions (*C4_Restrictions on gatherings*) with levels: 0 - no restrictions, 1 - restrictions on very large gatherings (the limit is above 1000 people), 2 - restrictions on gatherings between 101-1000 people, 3 - restrictions on gatherings between 11-100 people, 4 - restrictions on gatherings of 10 people or less.
Public transport closures (*C5_Close public transport*) with levels: 0 - no measures, 1 - recommend closing (or significantly reduce volume/route/means of transport available), 2 - require closing (or prohibit most citizens from using it). As from 13/3/2020 the value of this variable was always one (and zero before that date) there were insufficient differences in the values of the variable throughout the data in time for this variable to have any effect it was removed from the analysis.
Stay at home requirements (*C6_Stay at home requirements*) record orders to "shelter-in-place" and otherwise confine to the home with levels: 0 - no measures, 1 - recommend not leaving house, 2 - require not leaving house with exceptions for daily exercise, grocery shopping, and 'essential' trips, 3 - require not leaving house with minimal exceptions (eg allowed to leave once a week, or only one person can leave at a time, etc).
Internal movement restrictions (*C7_Restrictions on internal movement*) records restrictions on internal movement between cities/regions with levels: 0 - no measures, 1 - recommend not to travel between regions/cities, 2 - internal movement restrictions in place.
International travel restrictions (*C8_International travel controls*) records policy for foreign travellers, not citizens with levels: 0 - no restrictions, 1 - screening arrivals, 2 - quarantine arrivals from some or all regions, 3 - ban arrivals from some regions, 4 - ban on all regions or total border closure.

*Economic interventions (E)*

Income support (*E1_Income support*) records if the government is providing direct cash payments to people who lose their jobs or cannot work. It only includes payments to firms if explicitly linked to payroll/salaries. Levels: 0 - no income support, 1 - government is replacing less than 50% of lost salary (or if a flat sum, it is less than 50% median salary), 2 - government is replacing 50% or more of lost salary (or if a flat sum, it is greater than 50% median salary).

Dept relief (*E2_Debt/contract relief*) records if the government is freezing financial obligations for households (eg stopping loan repayments, preventing services like water from stopping, or banning evictions) with levels: 0 - no debt/contract relief, 1 - narrow relief, specific to one kind of contract, 2 - broad debt/contract relief.

Fiscal support measures (*E3_Fiscal measures*) includes the announced economic stimulus of spending. It only records amounts additional to previously announced spending in USD. It thus records monetary value in USD of fiscal stimuli and includes only any spending or tax cuts not included in E4, H4 or H5. This variable is recorded as daily time series but it is a continues monetary variable (i.e scale variable not ordinal factor) with values representing US dollars, and zero representing no new spending that day.

International financial support (*E4_International support*) includes the announced offers of Covid-19 related aid spending to other countries. It only records additional amounts to previously announced spending in USD. The variable is recorded as a continues monetary variable (scale variable not ordinal factor) daily time series. The variable was excluded from the analysis as it contained zero values exclusively.

*Health system policies (H)*

Public information campaigns (*H1_Public information campaigns*) records presence of public info campaigns with levels: 0 - no Covid-19 public information campaign, 1 - public officials urging caution about Covid-19, 2- coordinated public information campaign (eg across traditional and social media). The variable was excluded from the analysis as it contained exclusively values of level two.

Testing policy (*H2_Testing policy*) records the government policy on who has access to testing. This records policies about testing for current infection (PCR tests) not testing for immunity (antibody test) with levels: 0 - no testing policy, 1 - only those who both (a) have symptoms AND (b) meet specific criteria (eg key workers, admitted to hospital, came into contact with a known case, returned from overseas), 2 - testing of anyone showing Covid-19 symptoms, 3 - open public testing (eg "drive through" testing available to asymptomatic people).

Contact tracing (*H3_Contact tracing*) records the government policy on contact tracing after a positive diagnosis with levels: 0 - no contact tracing, 1 - limited contact tracing; not done for all cases, 2 - comprehensive contact tracing; done for all identified cases.

Emergency investment in healthcare (*H4_Emergency investment in healthcare*) includes the announced short term spending on healthcare system, eg hospitals, masks, etc. It only records the additional amount to previously announced spending per day in USD (scale variable not ordinal factor). As there was only one date with non-zero values (6/5/2020 with a value of 2,681,004 USD) there were insufficient levels of this variable to be included in the analysis.

Vaccine investment (*H5_Investment in vaccines*) includes the announced public spending on Covid-19 vaccine development. It records the additional amount to previously announced spending daily in USD (scale variable not ordinal factor). As there was only one date with non-zero values (4/5/2020 with a value of 3,242,505 USD) there were insufficient levels of this variable to be included in the analysis.

Facial coverings (*H6_Facial Coverings*) records policies on the use of facial coverings outside home with levels: 0 - No policy, 1 - Recommended, 2 - Required in some specified shared/public spaces outside the home with other people present, or some situations when social distancing not possible, 3 - Required in all shared/public spaces outside the home with other people present or all situations when social distancing not possible, 4 - Required outside the home at all times regardless of location or presence of other people.

Vaccination policy (*H7_Vaccination Policy*) records policies for vaccine delivery for different groups with levels: 0 - No availability, 1 - Availability for one of following: key workers/ clinically vulnerable groups / elderly groups, 2 - Availability for two of following: key workers/ clinically vulnerable groups / elderly groups, 3 - Availability for all of following: key workers/ clinically vulnerable groups / elderly groups, 4 - Availability for all three plus partial additional availability (select broad groups/ages), 5 - Universal availability.

*Epidemiological situation the country*

In an effort to compare the actual epidemiological situation in Greece against the web search effort, the daily score of $R_t$ was used. The $R_t$ is the average number of secondary infections produced when one infected individual is introduced into a susceptible host population (Anderson and May 1992). In practice, though not perfect, it indicates the average number of people who will contract a contagious disease from one person with that disease (Pandit 2020). The effective reproduction number, $R_t$, is defined as:

$$\mathcal{R}_t = \mathcal{R}_0^{(t)} \times (S_{t-1}/N),  \qquad [1]$$

where S is the number of susceptible individuals in a total population of N individuals, while $R^{(t)}_0$ is the average number of individuals infected by a single infected individual when everyone else is susceptible (Arroyo-Marioli et al. 2021). In practice $R_t > 1$ indicates a growing disease spread while $R_t < 1$ a disease spread that is diminishing, with larger values indicating higher spread. $R_t$ was calculated per day as part of the 'Our world in data' (Dong et al. 2020, Roser et al. 2020) epidemiological dataset publicly available at: (Our_World_in_Data 2021b). Daily values of $R_t$ were used averaged per week to match the temporal resolution and dates of web search effort data.

*Analysis*

*Artificial Neural Networks (ANN)*

ANN were used to quantify the complex relationship between web search effort (dependent variable) and the 16 explanatory variables. ANN are machine learning computing algorithms that can solve complex problems imitating animal brain in a simplified manner and can handle correlated independent variables (Rojas 1996, Hasson et al. 2020). Perception-type neural networks consist of artificial neurons or nodes, which is information processing units arranged in layers and interconnected by synaptic weights (connections); (Rojas 1996, Hasson et al. 2020). Neurons can filter and transmit information in a supervised fashion in order to build predictive model that clarifies data stored in memory (Rojas 1996, Hasson et al. 2020).

The Multilayer Perceptron (MLP) module was used to build the ANN and test its accuracy (Salgado et al. 2020). The MLP in ANN was trained with a back-propagation learning algorithm which uses the gradient descent to update the weights towards minimizing the error function (Salgado et al.

2020). The data were randomly assigned to 60% training, and 40% testing subsets. The training dataset is used to build the ANN model (Rojas 1996). The testing data is used to find errors and validate the model. Before training, all covariates were normalized using the formula (x−min)/(max−min), which returns values between 0 and 1.

For the hidden layer the hyperbolic tangent was used as activation function (Zamanlooy and Mirhassani 2013). The activation function Oj for each neuron of the *j*th output neuron takes real numbers as arguments and returns real values between -1 and 1. For the output layer, the identity function was used as activation function. Gradient descent optimization with the batch algorithm was used. The batch algorithm uses all records in the training dataset to update the synaptic weights between neurons (IBM 2016). The scaled conjugate gradient method was used for the batch training of the ANN (Marwala 2010). Before each iteration, the synaptic weights in the training dataset are updated. The algorithm finds the global error minimum by minimizing the total error made in the previous iteration (Møller 1993, IBM 2016).

Four parameters - initial lambda, initial sigma and interval center and interval offset - determine the way the scaled conjugate gradient algorithm builds the model. Lambda controls if the Hessian matrix is negative definite (Møller 1993). Sigma controls the size of weight change that affects the estimation of Hessian through the first order derivatives of error function (Rojas 1996). The parameters interval center $a_O$ and *a* force the simulated annealing algorithm to generates random weights that iteratively minimize the error function (IBM 2016). Initial lambda was set to 0.0000005, initial sigma to 0.00005. Interval center was defined as 0 and interval offset was set to ±0.5 (IBM 2016).

*Variable importance in ANN*

Variable importance was quantified using the outputs of the trained ANN model, in order to evaluate the effect of each input variable on the web search effort by using the variance based method (de Sá 2019, Ju et al. 2019). The input variables are ranked according to the sensitivity formula defined as:

$$S_i = \frac{V_i}{V(Y)} = \frac{V(E(Y|X_i))}{V(Y)},$$

[2]

where V(Y) is the unconditional output variance, E is the integral over Y|$X_i$, while the variance operator V implies a further integral over Xi. Variable importance is then computed as the normalized sensitivity. Si is the appropriate measure of sensitivity to rank the variables in order of importance for any combination of interactions and non-orthogonality among variables (Ju et al. 2019). The total sum of the overall V(I) of the ANNs is 1 (Ju et al. 2019).

*Temporal autocorrelation*

The correlation of web search in time was calculated, as this can provide information whether a high search effort week is likely to be followed by another high search effort one, or a low search effort is followed by a low search effort one, or if the search effort of a week is not indicative of the search effort of the following weeks (Moustakas and Evans 2017). To do so the temporal autocorrelation function for web search effort per week as a time unit lag was calculated (Reynolds and Madden 1988).

*Comparing web search effort and the epidemiological situation*

Inquiry-Charts, *I-charts* also termed as Shewart charts; (Ryan 2011, Montgomery 2020), were computed. An I-chart is a type of control chart used to monitor the process mean when measuring individuals at regular intervals from a process (Ryan 2011, Montgomery 2020). Each point on the chart represents the value of an individual observation (Ryan 2011, Montgomery 2020). The center line is the process mean (the mean of the individual observations). The control limits (upper and lower confidence intervals) are a multiple (k) of three sigma (k=3 σ) above and below the center line (Ryan 2011, Montgomery 2020). The process sigma is the standard deviation of the individual observations (Ryan 2011, Montgomery 2020). I-charts display individual data points and monitor mean and shifts in the process when the data points collected at regular intervals of time (Montgomery 2020). I-Charts may facilitate identifying the common and assignable causes in the process, if any (Rigdon et al. 1994). The green line on each chart represents the mean, while the red lines show the upper and lower control limits. An in-control process shows only random variation within the control limits (Ryan 2011). An out-of-control process has unusual variation, which may be due to the presence of special causes (Montgomery 2020). I-charts of web search effort, $R_t$, new cases, and new deaths were calculated.

**Results**

The trained ANN model with the 16 epidemiological and intervention variables exhibited a relative error of 0.058 (Sum of Squares error = 0.978) in the training dataset and a relative error of 0.237 in the test data (Sum of Squares error = 1.130). All error computations are based on the testing sample. The predictive accuracy of the trained ANN was $R^2$=91%, p<<0.001 between actual weekly values of web search effort of Covid-19 in Greece and the ANN model outputs (Fig. 1a). Model residuals exhibited a minor negative trend against the predicted value (Fig. 1b) but this deviance could not be differentiated from zero (confidence intervals where always crossing the zero line, $R^2$=5%, p<0.01).

Results regarding variable importance indicate that web search effort on Covid-19 in Greece was primarily featured by new deaths per week (relative importance of 100%), followed by international travelling restriction measures applicable to non-Greeks with relative importance of 95%, while new infections were ranked third with feature importance relative to other variables of 82% (Fig. 2). Testing policy was ranked fourth while restrictions on internal movement fifth, both with relative importance of 77 and 76% respectively (Fig. 2). School closing exhibited a relative importance of 71%, facial covering requirements 67%, staying at home requirements 66%, while restrictions on social gatherings 59% (Fig. 2). Work place closures featured a relative importance of 56%, cancelling public events 56%, fiscal measures 55%, while dept/contract relief 54% (Fig. 2). Ultimately, vaccination policy had a relative importance of 34%, income support 30%, and contact tracing 12% (Fig. 2).

The temporal autocorrelation of web search effort indicated no significant time lags between weeks indicating that the search effort of one week is not exhibiting significant correlation with the search effort of the next or previous weeks (Fig. 3).

The I-chart of web search effort indicated a negative deviation from confidence intervals during the first week in the analysis (week 1, week ending in 1 March 2020) as well as during week 14 (31 May 2020), while exceeded the upper confidence interval for two weeks in a row during weeks 24 & 25 (2 August 2020 to including the week starting at 16 August 2020), as well as also exceeding the upper confidence intervals during weeks 34 to 36 (18 October 2020 to 1 November 2020; Fig. 4a).

The I-chart of $R_t$ indicated that values were above the upper confidence interval during the first two weeks of available data (Fig. 4b; weeks 3 & 4; the $R_t$ needs some prior data before it can be calculated and thus may not be calculated during the first week that a case was found). Values of $R_t$ were below the lower confidence interval between weeks 8 to 13 (19 April 2020 to including the week starting at 24 May 2020), above the confidence interval the weeks 23 to 25 (2 August 2020 to including the week starting at 16 August 2020), and above the upper confidence interval the weeks 35 to 37 (25 October 2020 to including 8 November 2020; Fig. 4b). Values of $R_t$ were below the lower confidence interval the during the weeks 41 - 45 (6 December 2020 to 3 January 2021; Fig. 4b).

The I-chart values of new cases were always below the lower confidence interval till week 24 (9 August 2020), within the confidence interval till week 35 (25 October 2020), and above the higher confidence interval until week 43 (20 December 2020), within the confidence interval between weeks 44 - 49 (27 December 2020 - 31 January 2021), and above the confidence interval for the remaining two weeks of data (Fig. 4c).

The I-chart values of new deaths were below the lower confidence interval till week 33 (11 October 2020, week 31 was marginally within the confidence interval; Fig. 4d), within the confidence interval till week 37 (8 November 2020), and above the higher confidence interval until week 47 (17 January 2021), and within the confidence interval for the remaining four weeks of data (Fig. 4d).

**Discussion**

Results derived here studying indicated that fear of, or interest in death was the variable with the highest explanatory power in predicting internet search effort, implying that the first concern is either to stay alive, or interest and compassion to the individuals who passed way (Du et al. 2020, Lu and Reis 2021). This interest was higher than any financial related variable, or personal freedom of movement and entertainment, or every day habits related e.g. with schools and children (Giannopoulou et al. 2021). Internet searches on Covid were able to track Covid-19 spread (Azad and Devi 2020). This work indicates that a trained and validated data-driven model may provide useful insights regarding societal endeavors of future measures (Gozzi et al. 2020, Kar and Dwivedi 2020).

Interestingly the second variable for predicting Covid-19 web search effort was the opening of international borders applicable to non-Greek citizens and thus a variable directly related with tourism (Mariolis et al. 2020, Sharun et al. 2020). Opening international borders for non-Greeks and therefore the potential income generated by tourism featured considerably higher than any financial aid or facilitation derived by the state, as both debt/contract relief and fiscal measures were amongst the lower variables for predicting web search effort. This is in spite of the financial crisis driven by Covid-19 partially due to obligatory closures in a wide range of enterprises such as restaurants, bars, gyms, aesthetic centers, hotels, etc (Biskanaki et al. 2020, Mariolis et al. 2020). Interest in opening the international borders to non-Greek nationals has a potential two-fold implication: individuals may be searching about which foreign nationals are eligible for visiting Greece and technicalities of arrivals regarding potential testing requirements from the point of view of tourism-related interests or from an epidemiological perspective of how safe opening the international borders was (Mariolis et al. 2020, Sharun et al. 2020). The analysis conducted here may not differentiate between the two potential groups.

Another plausible option is that tourists that visited Greece during those dates (the peak season is 1-15 August) were extensively searching on Covid-19 to learn about the situation back in their home countries or to investigate the current situation in Greece. Foreign nationals searching for Covid-19 while in Greece are recorded within the Google the web search effort for Greece so long as they used a Greek IP address or connecting/roaming through a Greek internet network provider (Anderegg and Goldsmith 2014). However, data on web search effort are normalised by the other searches during those dates (see methods) and thus simply increasing the number of people searching would not result by default in higher web search effort on the topic (Ficetola 2013, McCallum and Bury 2014, Burivalova et al. 2018); the interest in the topic increased. It is also interesting to note that web search effort was not temporally correlated between weeks and thus there is no evidence that individuals are searching on the topic by habit or fear of the situation during the past few weeks, but rather dynamically tuning with the current weekly situation. Web search effort in Greece did not temporally match the one recorded in Cyprus during the research study period (Anastasiou et al. 2021).

It is a fact that $R_t$ was peaking in Greece during the peak of the tourist season and when international borders were not restricted. Greece was also topping the list of tourist-imported Covid-19 cases in the UK followed up the 2020 summer season among all other examined countries (Aggarwal et al. 2021). During the first half of August 2020, web search effort around Covid-19 is also peaking despite the fact that new deaths or new cases were not peaking. It was however widely discussed whether the opening of international borders to tourism with low testing frequencies or without a negative PCR Covid-19 test of incoming individuals was a safe practice (Pavli et al. 2020, Rocklöv et al. 2020, Sharun et al. 2020). Thus, the interest in the opening of international borders may have an interaction effect with the interest in testing policy in the total web search effort. The temporal coincidence of web search effort peak with the peak of $R_t$ is interesting given that the $R_t$ is not something regularly reported in the news, media, or non-scientific websites as more often than not new cases, or deaths, or intervention measures are reported (Liu et al. 2020). Web search effort peak was also coinciding with $R_t$ peaks in October, from 27 October 2020, the date when new cases exceeded the threshold of 1,000 per day. Thus findings reported here confirm both that web search effort can be used to predict peaks in COVID-19, as well as that peaks of online searches precede the reported confirmed cases and deaths (Lampos et al. 2021).

Testing policy appeared as the fourth most predictive variable for predicting web search effort and above other variables such as exiting home restrictions, restrictions on gatherings, movements in other regions in Greece, obligatory use of masks, or school closures. This may seem as a surprising result given that testing policy is not affecting or restricting everyday life. However, testing frequency is directly related both with disease spread control through quarantining infected individuals (Wells et al. 2021). In addition, Greece has been reported not to test as frequent as other countries as exhibited by a high ratio of deaths per cases in comparison to other countries (Fouda et al. 2020). Greece and people could potentially worry regarding new cases as these could be either under-reported or disease spread via untested individuals. Frequent testing could control the disease and reduce the duration of quarantine (Peto 2020, Wells et al. 2021) and it is known to de-synchronize disease spread in other diseases too (Moustakas et al. 2018).

The idea that crowd wisdom might reflect the reality better than expert opinion or any single individual, has been considered as provocative in the past (Galton 1907, Prelec et al. 2017). Web search has the cultural, geographical, social, temporal diversity and a sample size that is hard to be ignored (Surowiecki 2005, Sunstein 2006). The way people feel or use the internet was narrowed down here by examining exclusively web search effort in terms of dates and underlying parameters and not what exactly do they search regarding the topic of Covid-19. The use of internet or reactions to Covid-19 in general is complex phenomenon depending on several other factors unaccounted for here, such as age, gender, geographic location, education and financial status (Parlapani et al. 2020, Kamenidou et al. 2021), fake news (Guess et al. 2018), social circles and networks (Galesic et al. 2018), and information gerrymandering (Stewart et al. 2019). Still web search behaviour regarding other diseases exhibits universality with web search effort largely coinciding with epidemiological parameters (Alicino et al. 2015). Still, in Greece web search effort peaked during the same time periods with $R_t$ indicating that web search is not necessarily dominated by such issues and reflected the reality better than the recorded situation at least in the scarcity of testing data.

Ultimately this study could not properly evaluate interventions that were recently implemented or in practice where not exercised by the government as due to insufficient factor level variation. In specific, vaccination policy was included in the analysis (Huang et al. 2021) but its efficacy on web search effort may be hindered by the small temporal sample size of its applicability. The Greek government financed only once in emergency investments in health care throughout the Covid-19 pandemic and to that end the (scarcity of) this intervention (Siettos et al. 2021) may not be quantified. Contact tracing is included in the analysis but ranked as the least important out of 16 variables for predicting web search effort; while there was sufficient variation in the values to be included in the analysis, in practice it was not an intervention exercised by the government to a high extend as done in other countries (Cho et al. 2020, Bradshaw et al. 2021). In addition this analysis cannot account for individuals in Greece briefed on Covid-19 exclusively by television or radio, or without internet access, or the ones using a search engine other than Google.

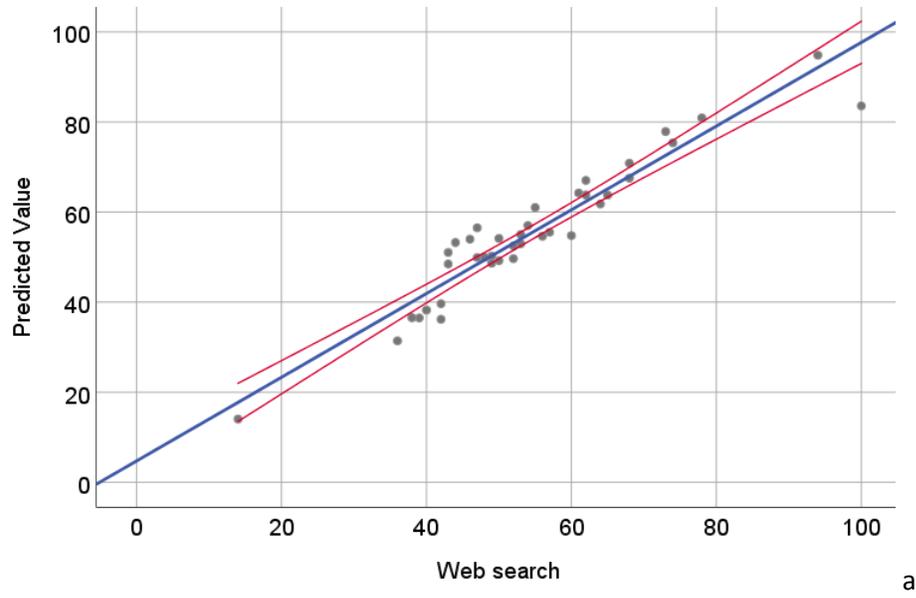
a

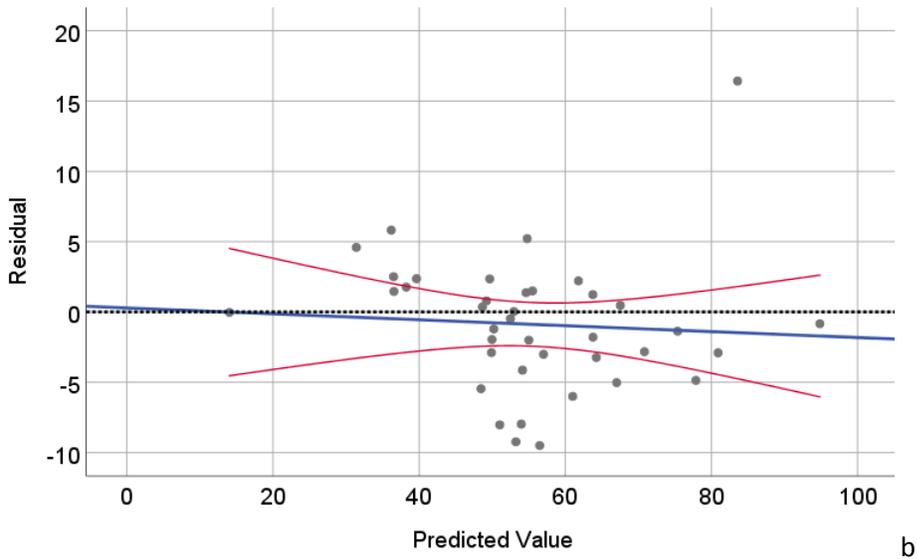
b

**Figure 1.** Accuracy of the Artificial Neural Network (ANN) model. The ANN model was trained with 75% of the data and tested against the 25% of the data. Error computations are based on the testing data. The blue line indicates the regression, while red lines indicate a 95% confidence interval of the regression. Actual values are plotted in grey circles. **a.** Linear regression between the trained ANN predicted values versus actual values of web search effort. **b.** Residuals of the regression between the trained ANN versus the predicted values. The horizontal dotted black line indicates zero value. The 95% confidence intervals crossing zero indicate that residuals may not be differentiated from zero although there is a weak negative correlation between residuals and predicted values.

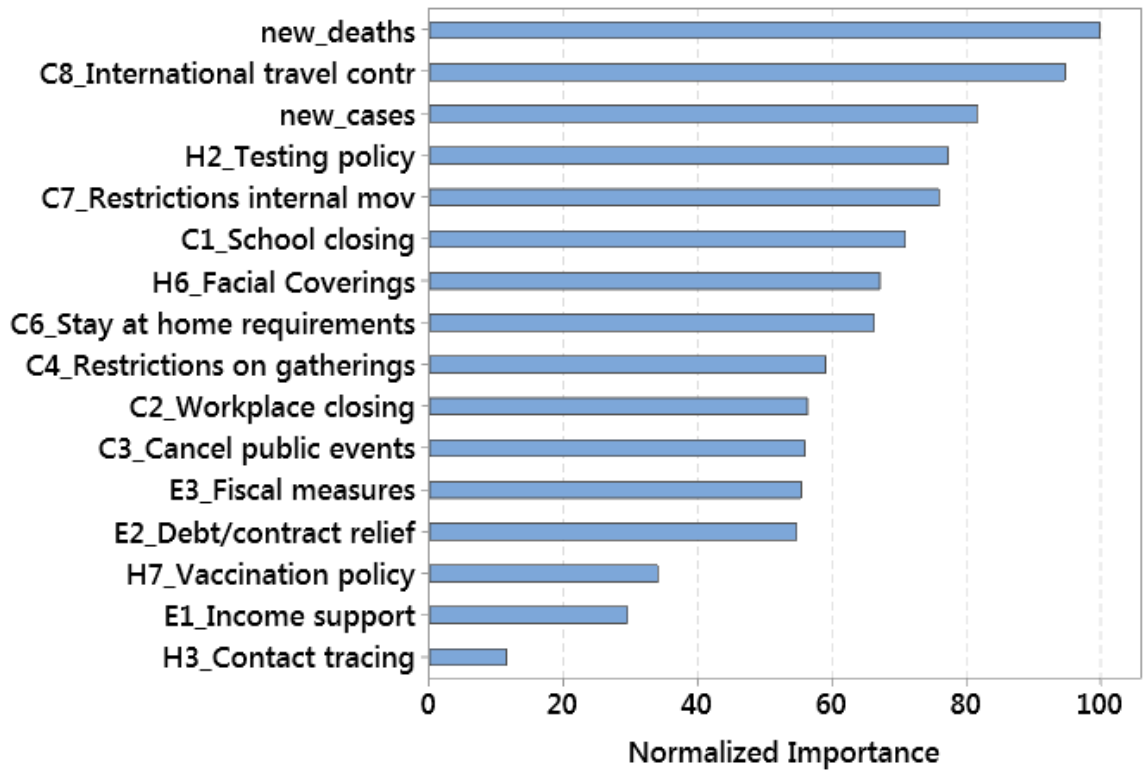

**Figure 2.** Normalised importance of the 16, *containment and closure policies (C)* health system policies (H), economic interventions (E), and epidemiological (new cases, new deaths) variables of the trained ANN model for predicting web search effort regarding Covid-19 as a topic in Greece from February 2020 to February 2021 on a weekly time scale.

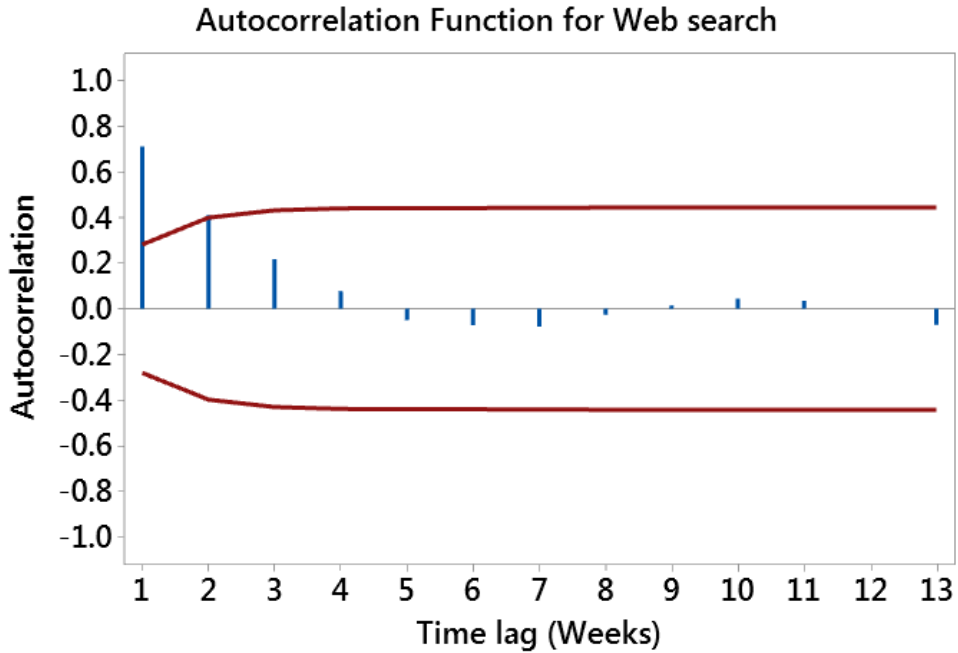

**Figure 3.** Temporal autocorrelation function of web search effort across weekly time lags. Values close to 1 or -1 indicate strong (positive or negative) correlation while values close to zero no correlation. The horizontal red lines indicate a 95% significance confidence intervals, while the vertical blue lines the actual correlation value for that time lag. Correlation values within the 95% confidence interval are not significant and they may not be differentiated from random correlations.

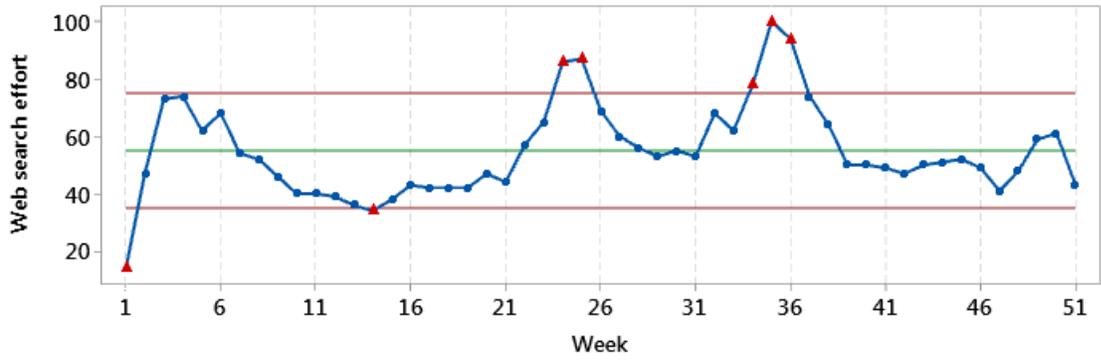

a

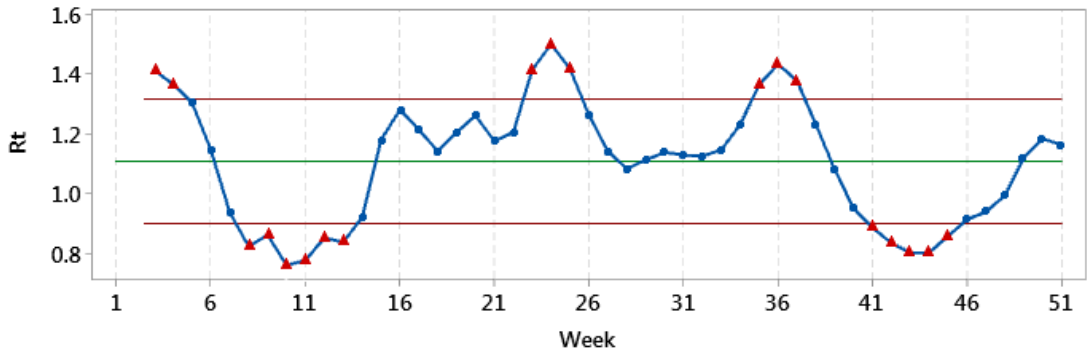

b

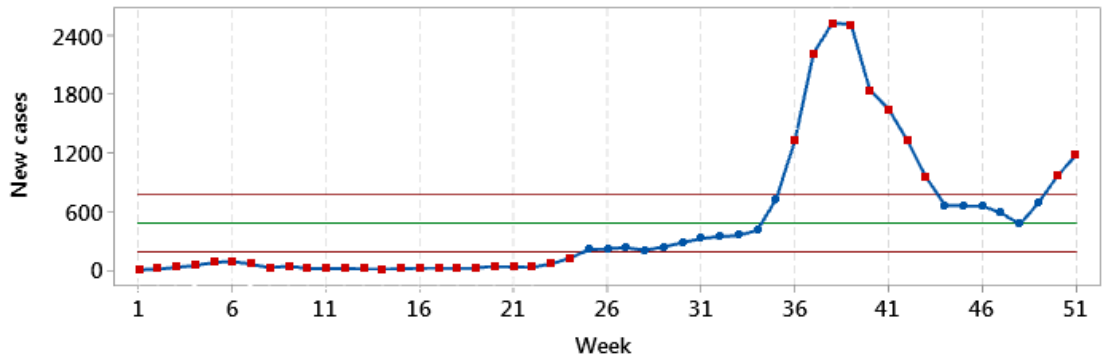

c

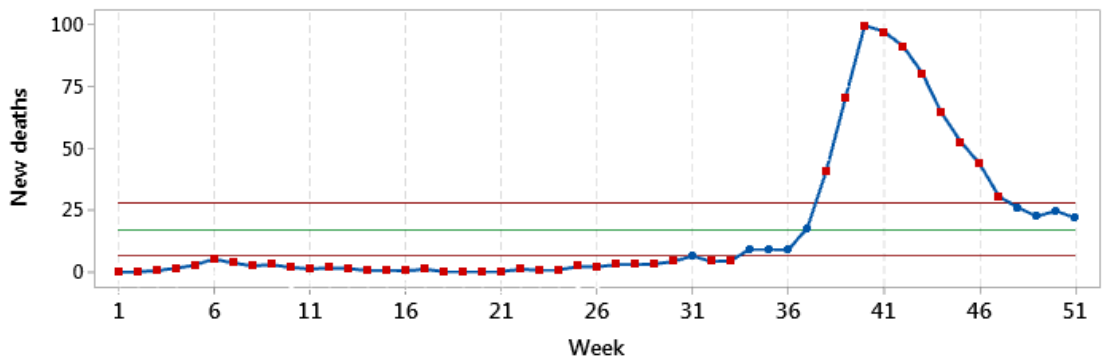

d

**Figure 4.** I-Charts of data in time on a weekly scale from 26 February 2020 (first case, week 1) to 14 February 2021 (week 51). Throughout the analysis week 1 is the week 24 Feb 2020 - 1 March 2020,

while week 51 is the week 8 Feb 2021 - 14 Feb 2021. The horizontal green line indicates the grand mean, while the uper and lower red lines indicate three sigma above and below the mean respectively. The blue line indicates the actual values during that week. All values deviating the confidence intervals are depicted in red symbols. **a.** I-chart of web search effort. **b.** I-chart of $R_t$. **c.** I-chart of new cases. **d.** I-chart of new deaths.